 \documentclass[aps,prb,twocolumn,showpacs,amsmath,amssymb]{revtex4}

\usepackage{graphicx}
\usepackage{dcolumn} 
\usepackage{bm}      
\usepackage{verbatim}%
\usepackage{epstopdf}

\begin{document}


\title
{
Monte Carlo simulations of the structure of Pt-based bimetallic nanoparticles
}


\author{ Kayoung Yun }
\affiliation
{
School of Advanced Materials Engineering, Kookmin University, Seoul 136-702, Korea
}

\author{ Jung Soo Oh }
\affiliation
{
Computational Science Center, Korea Institute of Science and Technology, Seoul 130-650, Korea
}
\author{ Jung-Hae Choi }
\affiliation
{
Computational Science Center, Korea Institute of Science and Technology, Seoul 130-650, Korea
}
\author{ Seung-Cheol Lee }
\email
{leesc@kist.re.kr}
\affiliation
{
Computational Science Center, Korea Institute of Science and Technology, Seoul 130-650, Korea
}

\author{ Yong-Hun Cho }
\affiliation
{
School of Advanced Materials Engineering, Kookmin University, Seoul 136-702, Korea
}
\author{ Pil-Ryung Cha }
\affiliation
{
School of Advanced Materials Engineering, Kookmin University, Seoul 136-702, Korea
}
\author{ Jaegab Lee }
\affiliation
{
School of Advanced Materials Engineering, Kookmin University, Seoul 136-702, Korea
}
\author{ Ho-Seok Nam }
\email
{hsnam@kookmin.ac.kr}
\affiliation
{
School of Advanced Materials Engineering, Kookmin University, Seoul 136-702, Korea
}

\date{\today}

\begin{abstract}
Pt-based bimetallic nanoparticles have attracted significant attention as a promising replacement for expensive Pt nanoparticles. In the systematic design of bimetallic nanoparticles, it is important to understand their preferred atomic structures. However, compared with unary systems, alloy nanoparticles present more structural complexity with various compositional configurations, such as mixed-alloy, core-shell, and multishell structures. In this paper, we developed a unified empirical potential model for various Pt-based binary alloys, such as Pd-Pt, Cu-Pt, Au-Pt, and Ag-Pt. Within this framework, we performed a series of Monte Carlo (MC) simulations that quantify the energetically favorable atomic arrangements of Pt-based alloy nanoparticles: an intermetallic compound structure for the Pd-Pt alloy, an onion-like multi-shell structure for the Cu-Pt alloy, and core-shell structures (Au@Pt and Ag@Pt) for the Au-Pt and Ag-Pt alloys. The equilibrium nanoparticle structures for the four alloy types were compared with each other, and the structural features can be interpreted by the interplay of their material properties, such as the surface energy and heat of formation. 
\end{abstract}

\pacs{61.46.+w, 36.40.Ei, 64.70.Nd}


\maketitle


\section{\label{sec:level_1intro} Introduction}

Due to their high surface-to-volume ratio or quantum effects at extremely small sizes, metal nanoparticles exhibit unique chemical and physical properties that are distinct from bulk materials~\cite{Ferrando:NanoclusterRMP}. Among the many applications of nanoparticles, such as electronic, optical, and  magnetic devices~\cite{Kruis:NPsReview, Johnston:MetalCluster, Ferrando:NanoalloysChemRev, Cortie:NPsPlasmonic}, their electro-catalytic effects are the most beneficial, especially for many effective energy conversion techniques, e.g., the performance of proton-exchange membrane fuel cells (PEMFCs) and direct methanol fuel cells (DMFCs) as power resources for electric devices~\cite{Chen:NPsPtCatalysts, Gasteiger:PtCatalystsPEMFCs, Strasser:NatureCuPt, Cho:NPsPdPtCatalystsPEMFCs}.

Platinum (Pt) is practically the only catalyst active for hydrogen oxidation, methanol oxidation and oxygen reduction at low temperature and it is one of the most efficient metal catalysts for fuel cells and similar energy conversion techniques~\cite{Gasteiger:PtCatalystsPEMFCs}. As mentioned in many articles~\cite{Chen:NPsPtCatalysts, Gasteiger:PtCatalystsPEMFCs, Strasser:NatureCuPt, Cho:NPsPdPtCatalystsPEMFCs}, Pt nanoparticles with sizes of a few nanometers provide exclusively outstanding catalytic activity/stability. The primary disadvantage of using Pt nanoparticles in applications is the high cost of Pt (which is one of the main obstacles to the commercialization of fuel cells). Two solutions have been suggested for reducing the industrial demand for Pt in fuel cells: replacing Pt with alternative non-noble catalysts and reducing the Pt-loading by exploiting non-precious supports~\cite{Gasteiger:PtCatalystsPEMFCs}. In this regard, the design of Pt-based bimetallic nanoparticles to replace Pt nanoparticles has attracted significant attention~\cite{Cho:NPsPdPtCatalystsPEMFCs, Yoo:NPsPdPtCatalystsPEMFCs, Strasser:NatureCuPt}. For the systematic design of bimetallic nanoparticle catalysts, it is very important to understand their stable atomic structures because the chemical and physical properties of various Pt-based alloy metal nanoparticles are highly dependent on their size, shape, composition, morphology, and structural stability.

Compared with unary catalysts, alloy catalysts present more structural complexity because the two components can exhibit various structural configurations/modifications. For example, bimetallic nanoparticles can exhibit the structures of ordered/random mixed-alloy~\cite{Ferrando:NanoalloysChemRev}, core-shell~\cite{Johnson:NPsCoreShellTrends, Yugang:NPsAuAg_CoreShell}, and multishell nanoparticles~\cite{Baletto:Onion, Cheng:Onion}. The equilibrium structure of bimetallic nanoparticles is also sensitive to the alloy components, involving the interplay of several material properties, such as surface energy, heat of mixing, and strain effect due to the atomic size difference. Because of these complex factors, it is difficult to predict the most stable structure of bimetallic nanoparticles and their thermodynamic properties. In this paper, we developed a unified embedded atom method (EAM) potential model for simulating the structures of various Pt-based binary alloys, such as Pd-Pt, Cu-Pt, Au-Pt, and Ag-Pt. Within the framework of our potential model, we performed a series of Monte Carlo (MC) simulations to quantify  the energetically favorable atomic structures of Pt-based alloy nanoparticles.

\section{\label{sec:level_2potential} Interatomic Potential Model}

\subsection{\label{sec:sublevel_21ElemPototential} Embedded atom method potentials for Pt, Pd, and novel metals (Ag, Au, Cu)}

The core ingredient of atomic-scale simulations is how to evaluate the potential energy of systems and the interatomic forces as a function of the positions of the atoms. Although it originates from the electronic structures, in classical atomistic simulations, such as molecular dynamics (MD) and Monte Carlo (MC) methods, the interaction between atoms is usually described by empirical potentials, and therefore the reliability of the simulation result is entirely dependent on the reality and accuracy of the interatomic potentials. For most metallic systems, the embedded atom method (EAM) potential is widely used, and several potentials have been developed for elemental Pt and some other face-centered cubic (fcc) alloys~\cite{Voter:EAMfccMetals, Foiles:EAMfccMetals, Johnson:EAMfccMetals, MeiDavenport:EAMfccMetals, Cai:EAMfccMetals}. Especially in many applications of alloy systems, EAM potential models of analytic functional form have been frequently used because of their simplicity and extendibility to multi-component systems. Here, we adopted EAM potential models for fcc metals that were originally developed by A. Voter~\cite{Voter:EAMfccMetals} and later modified by us~\cite{Nam:TuningTm}. The total energy of the system is given by the usual EAM form: 
\begin{equation}
\label{eq:EAM_TotalEnergy}
E ~ = ~ \sum_{i} \left[ F_{s_i} ( \bar \rho_i )
~ + ~ \frac {1} {2} \sum_{j \neq i} \phi_{s_i \operatorname{-} s_j} ( r_{ij} ) \right], 
\end{equation}
where $ F_{s_i} ( \rho )$ is the energy associated with embedding atom of type $s_i$ in a uniform electron gas of density $\rho$ and $\phi_{s_i \operatorname{-} s_j} ( r )$ is a pairwise interaction between atoms of type $s_i$ and $s_j$ separated by a distance $r$. 
The electron density is given by
\begin{equation}
\label{eq:EAM_ElectronDensity}
\bar \rho_i = \sum_{j(\ne i)} f ( r_{ij} ), 
\end{equation}
where the atomic electron density function $f(r)$ is taken as the density of a hydrogenic $4s$ orbital:
\begin{equation}
\label{eq:EAM_DensityFitting}
f(r) = f_{0}~r^{6} \left( e^{- \beta r} + 2^9 e^{ -2 \beta r} \right).  
\end{equation}
Here, $\beta$ is an adjustable fitting parameter that quantifies the distance over which the electron density decays away from an atom position and $f_0$ is a prefactor. (It can be an arbitrary value for a unary system because it cancels out when combined with the embedding function, but it is an adjustable parameter that should be determined for an alloy system.) We chose $f_0$ to be $1/(f(r_{eq}N_{1st})$ for convenience, so that the electron density at the equilibrium crystals is approximately normalized to unit value.

Within the formalism of A. Voter~\cite{Voter:EAMfccMetals}, the pair potential term, $\phi(r)$, is chosen to take a Morse potential form with a minor additional term:
\begin{eqnarray}
\label{eq:EAM_PairPotential}
\phi (r) = 
&& - D_{M} \left[ 2 e^{- \alpha_{M} \left( r - R_{M} \right)} - e^{- 2\alpha_{M} \left( r - R_{M} \right)} \right]
\nonumber\\
&& + \frac{64 \delta}{(r_3 - r_2)^6} (r-r_2)^3 (r_3-r)^3
\nonumber\\
&& \; \; \times \theta (r-r_2) \theta(r_3-r),
\end{eqnarray}
where $D_{M}$, $R_{M}$, and $\alpha_{M}$ are adjustable fitting parameters that define the well-depth of the Morse pair-interaction, the position of the minimum, and a measure of the curvature at the minimum, respectively. The second term was introduced to tune the melting point of the EAM model: the parameter $\delta$ represents the magnitude of the pair-interaction tuning between the second and third nearest neighbor positions, $r_2$ and $r_3$, with the aid of the Heaviside step-function,  $\theta (r)$. Details regarding the meaning of this term will be provided elsewhere~\cite{Nam:TuningTm}.

Finally, the embedding function was numerically determined so that the total energy of the reference system as a function of dilation satisfies the following universal binding energy relation~\cite{Rose:EAMUniversalFeatures}: 
\begin{equation}
\label{eq:EAM_UniveralRelation}
E(a) = - E_{0} \left[ 1 + \alpha \left( \frac{a}{a_{0}} - 1 \right) \right]
\exp \left[ - \alpha \left( \frac{a}{a_{0}} - 1 \right) \right], 
\end{equation}
with $\alpha = \sqrt{ 9B\Omega / E_0} $, where $a$ is the dilated lattice constant, $a_{0}$ is the equilibrium lattice constant and $E_{0}$, $B$, and $\Omega$ are the cohesive energy, bulk modulus, and equilibrium atomic volume of the reference lattice, respectively. 
The potential interactions were smoothly cut off at $r=r_{cut}$ (usually between the third- and fourth-nearest-neighbor shells of a static fcc crystal) to ensure that the interatomic potential and its first derivatives are continuous.

Because it takes such an analytic functional form, this potential automatically gives exact matching with the experimental values of the equilibrium lattice parameter $a_0$, cohesive energy $E_0$, and bulk modulus $B$ via Eq.~(\ref{eq:EAM_UniveralRelation}). Thus, we need to fit the remaining six adjustable parameters ($D_M, R_M, \alpha_M, \beta, \delta$, and $r_{cut}$) for a single component, and those are obtained by optimization relative to some material properties. The parameters in the Voter's potentials were fitted to a series of basic properties, including the three cubic elastic constants ($C_{11}$, $C_{12}$, and $C_{44}$) and the vacancy formation energy, $E_{vac}$, of the face-centered cubic (fcc) crystal for Pt, Pd, and the novel metals~\cite{Simmons:SingleCrystal}. Here, instead of using the original parameter set suggested by Voter~\cite{Voter:EAMfccMetals}, we re-optimized the potential parameters (except $r_{cut}$, which is of less importance), by including the surface energy in the target properties~\cite{Nam:TuningTm}. The resulting parameter sets are shown in Table~\ref{tab:table1} and the calculated material properties are compared with the target values in Table~\ref{tab:table2}. 

\begin{figure}[!tbp]
\includegraphics[width=0.40\textwidth]{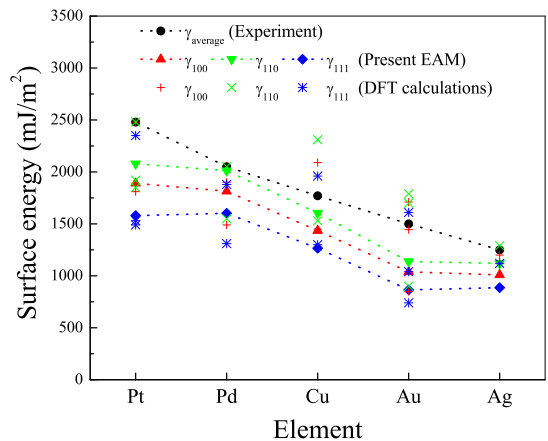}
\caption
{\label{fig:SurfaceE}
Surface energies of the five metals: Pt, Pd, Cu, Au, and Ag. The calculated values based on our EAM model are compared with the first-principle calculations and the experimental data. The experimental surface energies (solid circles) correspond to averaged values over the various surfaces, which are from Refs.~\onlinecite{Boer:CohesionInMetals} and \onlinecite{Tyson:EsurfExp}.
The calculated values are measured on the $(100)$, $(110)$, and $(111)$ surfaces (Refs.~\onlinecite{Skriver:EsurfLDA, Marzari:EsurfPdPtAuGGAPBE, Entel:EsurfPtGGAPBE, Gross:EsurfCuLDA}). The first-principle calculations show diverse data depending on their exchange-correlation functionals.
}
\end{figure}

Among the various materials properties, we believe that the surface energies of the elements are of particular importance for the simulation of nanoparticles. Figure~\ref{fig:SurfaceE} shows the surface energies of the five elements (Pt, Pd, Cu, Au, and Ag) that were calculated by our EAM potential compared with first-principle calculations~\cite{Skriver:EsurfLDA, Marzari:EsurfPdPtAuGGAPBE, Entel:EsurfPtGGAPBE, Gross:EsurfCuLDA} and experimental data~\cite{Boer:CohesionInMetals, Tyson:EsurfExp}. While there is a diverse data distribution in the first-principle calculation results, the order of the experimental surface energy was consistently $\gamma_{Pt} > \gamma _{Pd} > \gamma _{Cu} > \gamma _{Au} > \gamma _{Ag}$. The surface energies of the atomic species are expected to be one of the most influential properties in the energetics of nanoparticles due to the high surface-to-volume ratio, and we believe that the relative value of the surface energy will play a role in determining the most favorable nanoparticle structures and their structural behavior. Note that the present EAM potential model predicts the correct order of the surface energies among the elements even better than the first-principle calculations and hence is expected to be transferable to nanoparticle systems with a large surface-to-volume ratio. 

\begin{figure}[!tbp]
\includegraphics[width=0.40\textwidth]{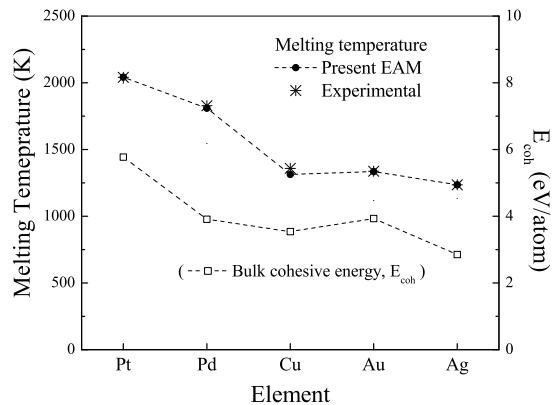}
\caption
{\label{fig:MeltingPt}
Melting temperatures of the five metals (Pt, Pd, Cu, Au, and Ag): Our EAM model reproduces the experimental melting temperatures to within less than 5\%.
}
\end{figure}

Figure~\ref{fig:MeltingPt} shows the melting temperature of the five elements predicted by our EAM potential model. (The melting points of the fcc metals were obtained via microcanonical ensemble molecular dynamics simulations of the solid-liquid coexistence~\cite{Morris:CoexistingMD}.) With the aid of the tuning term in the pair-potential function, our EAM model reproduces the experimental melting temperature within less than a 5\% difference. 

\begin{table}[!tbp]
\caption
{\label{tab:table1}
Fitted potential parameters for five fcc metals: Pt, Pd, Cu, Au, and Ag. 
}
\begin{ruledtabular}
\begin{tabular}{cccccc}
Parameter   & Pt & Pd & Cu & Au & Ag \\
\hline \\  [-1.5ex]
$\beta$ (\AA$^{-1}$)    & 3.580 & 3.400 & 3.990 & 3.795 & 3.960  \\  [0.5ex]
$\alpha_M$ (\AA$^{-1}$) & 1.779 & 1.501 & 1.838 & 1.753 & 1.685  \\  [0.5ex]
$D_M$ (eV)              & 0.783 & 1.682 & 0.781 & 0.695 & 0.693  \\  [0.5ex]
$R_M$ (\AA)             & 2.577 & 2.343 & 2.321 & 2.537 & 2.538  \\  [0.5ex]
$r_{cut}$ (\AA)         & 5.576 & 5.412 & 4.961 & 5.516 & 5.542  \\  [0.5ex]
$\delta$ (eV)           & 0.01  & 0.06  & 0.0   & 0.049 & 0.0235 \\  [0.5ex]
\end{tabular}
\end{ruledtabular}
\end{table}

\begin{table}[!tbp]
\caption
{\label{tab:table2}
Materials properties used in fitting for fcc Pt, Pd, Cu, Au, and Ag. Where two numbers are given, the top number is the calculated value predicted by the present EAM model, and the lower number in parentheses is the experimental value from Ref.~\onlinecite{Simmons:SingleCrystal} (elastic properties) or the first-principle calculation results from Refs.~\onlinecite{Skriver:EsurfLDA, Marzari:EsurfPdPtAuGGAPBE, Entel:EsurfPtGGAPBE, Gross:EsurfCuLDA} (surface energies). 
}
\begin{ruledtabular}
\begin{tabular}{cccccc}
Properties   & Pt & Pd & Cu & Au & Ag \\
\hline \\  [-1.5ex]
$a_0$ (\AA)         & 3.924 & 3.891 & 3.615 & 4.078 & 4.085  \\ [1.0ex]
$E_{coh}$ (eV/atom) & 5.77  & 3.91  & 3.54  & 3.93  & 2.85   \\ [1.0ex]
$B$ (GPa)           &  283  &  195  &  142  &  167  &  104   \\ [1.0ex]
$C_{11}$ (GPa)      &  319  &  234  &  180  &  185  &  121   \\
                    & (347) & (234) & (176) & (186) & (124)  \\ [1.0ex]
$C_{12}$ (GPa)      &  264  &  175  &  124  &  158  &  94.9  \\
                    & (251) & (176) & (125) & (157) & (93.4) \\ [1.0ex]
$C_{44}$ (GPa)      &   77  &  70.5 &  79.8 &  40.3 &  43.7  \\
                    &  (77) & (71.2)& (81.8)& (42.0)& (46.1) \\ [1.0ex]
$E_{v}$ (eV)        &  1.4  &  1.58 &   1.3 &   1.0 &   1.2  \\
                    & (1.5) & (1.54)&  (1.3)&  (0.9)&  (1.1) \\ [1.0ex]
$\gamma_{100}$ (mJ/m$^2$)  & 1891 & 1815 & 1435 & 1038 & 1009   \\
                           &(1889)&(1900)&(1450)&(1444)&(1200)  \\ [1.0ex]
$\gamma_{110}$ (mJ/m$^2$)  & 2079 & 2012 & 1605 & 1138 & 1120   \\
                           &(1920)&      &(1530)&(1700)&(1290)  \\ [1.0ex]
$\gamma_{111}$ (mJ/m$^2$)  & 1580 & 1602 & 1266 &  865 &  886   \\
                           &(1531)&(1880)&(1300)&(1040)&(1120)  \\ [1.0ex]
\end{tabular}
\end{ruledtabular}
\end{table}

\subsection{\label{sec:sublevel_22AlloyPotential} Interatomic potentials for Pt-based alloys}

Significant effort has also been made to develop empirical potential models for metallic alloys. For alloy potentials, we describe the interaction between atoms of different species, which are often called the cross-interaction potentials. In the early versions of EAM potentials taking analytic forms, cross-interaction potentials were proposed to take a simplified combination of the individual pair-interactions based on the potentials for pure metals, e.g., the cross-interaction potential was assumed to be the geometric mean of the monatomic pair potentials~\cite{Foiles:EAMfccMetals} or to be a density-weighted combination of the monatomic pair potentials~\cite{Johnson:EAMfccMetals, Cai:EAMfccMetals}. This approach was fairly reasonable for several binary solid solutions of fcc metals, but not sufficient for a quantitative study of the alloy properties. Currently, even for a system in which no adequate experimental data are available, first-principles calculations provide the necessary data for fitting the cross-interaction potentials, such as the formation enthalpy and the lattice constant of the alloys, and, as a result, more accurately parameterized potentials can be constructed for individual alloys. However, we are rarely aware of the development of potentials that were optimized to reproduce the properties of various binary Pt-based alloy systems. In this study, we optimized the adjustable parameters for the cross-interaction  based on both the formation enthalpy and the lattice constant of the alloys. (For cross-interaction, $\delta$ was set to zero.) The determined potential parameters for four Pt-based alloys are shown Table~\ref{tab:table3}. 

\begin{table}[!tbp]
\caption
{\label{tab:table3}
Parameters for the Pt-based alloy potential.
}
\begin{ruledtabular}
\begin{tabular}{ccccc}
Parameter   & Pt-Pd & Pt-Cu & Pt-Ag & Pt-Au \\
\hline \\  [-1.5ex]
$\alpha_M$ (\AA$^{-1}$) & 1.701 & 1.734 & 1.820 & 1.588  \\  [0.5ex]
$D_M$ (eV)              & 1.259 & 0.770 & 0.760 & 0.792  \\  [0.5ex]
$R_M$ (\AA)             & 2.448 & 2.517 & 2.555 & 2.430  \\  [0.5ex]
\end{tabular}
\end{ruledtabular}
\end{table}

\begin{figure}[!tbp]
\includegraphics[width=0.36\textwidth]{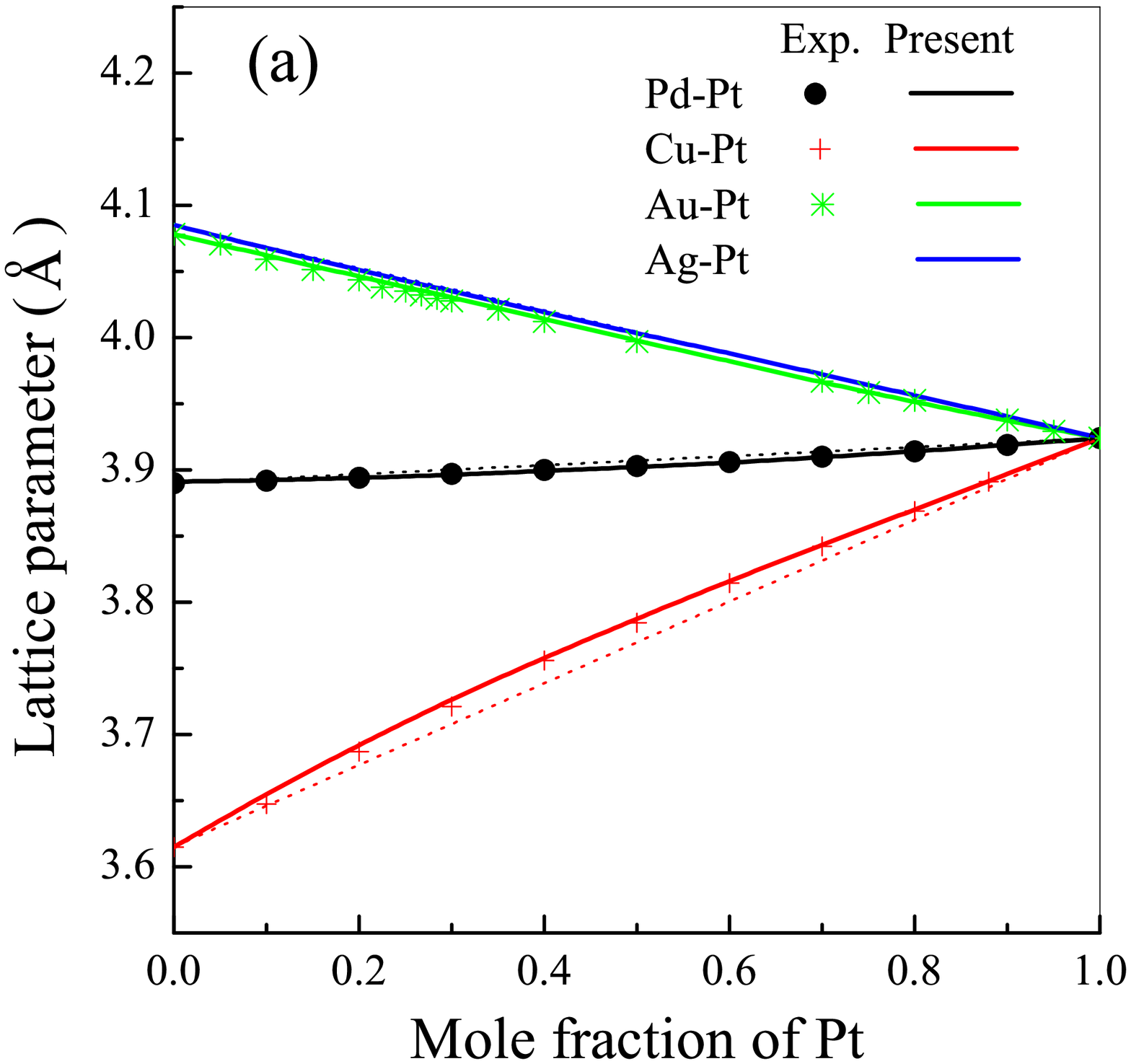}
\includegraphics[width=0.36\textwidth]{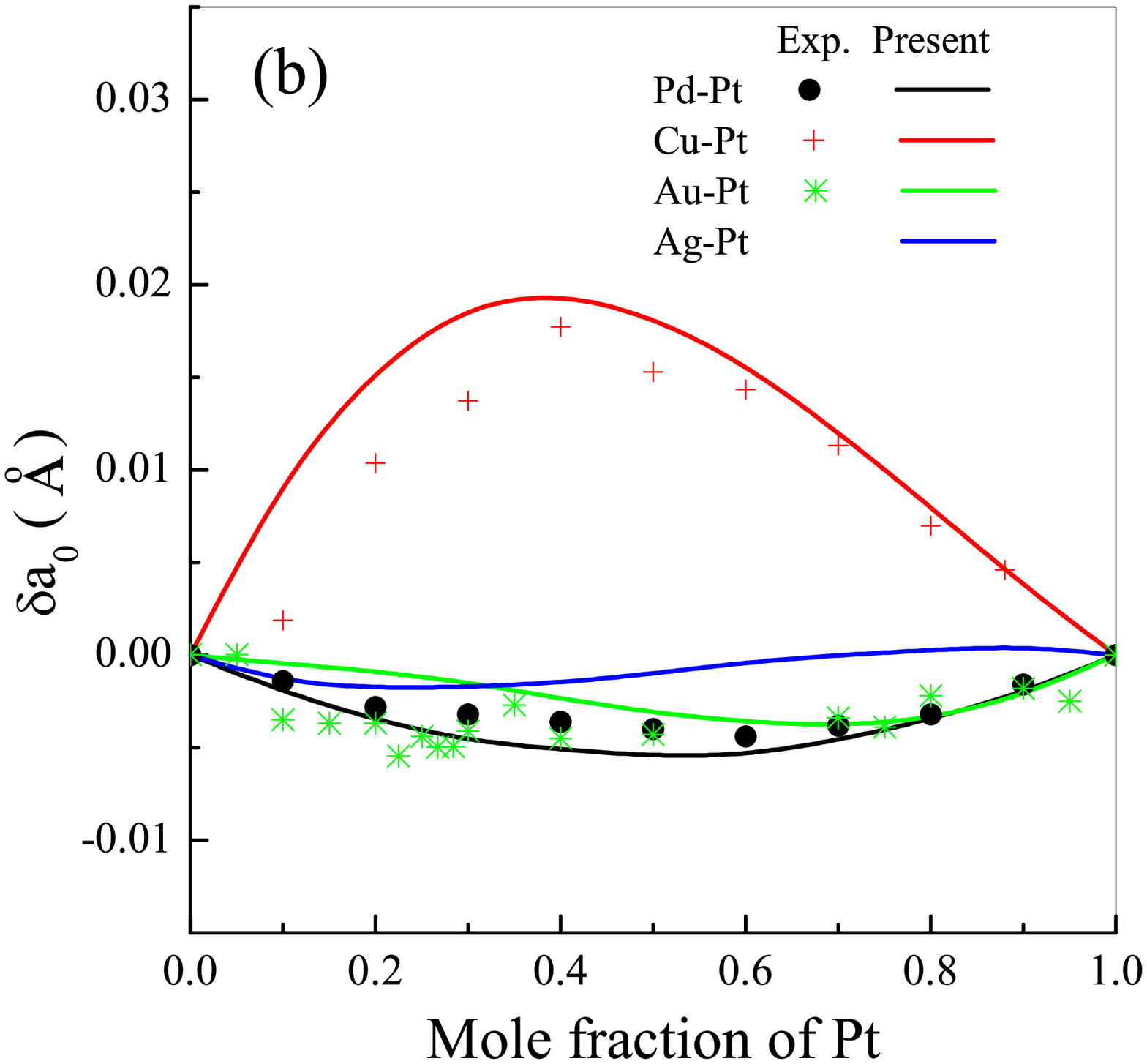}

\caption
{ \label{fig:latticepara}
(a) Lattice parameters and (b) their variation of the departure from Vegard's law for the Pd-Pt, Cu-Pt, Ag-Pt, and Au-Pt bulk systems as a function of the Pt concentration. The lines are the results of the present EAM model, and the solid symbols represent the experimental data. 
}
\end{figure}

Figure~\ref{fig:latticepara} shows the equilibrium lattice constants of four Pt-based alloys as a function of the Pt mole fraction together with the published experimental data~\cite{Pearson:HandbookLattice, Abe:ThermoCuPt, Karakaya:ThermoAuPt, Durussel:ThermoAgPt}. (For Ag-Pt, we could not find experimental data on the mixing enthalpy.) Because the Ag-Pt system forms an intermetallic compound, experimental data were not compared for Ag-Pt. In the case of Pd-Pt, Cu-Pt, and Au-Pt, the present model showed reasonable agreement with the experimental results. Note that the Cu-Pt system showed somewhat positive deviation from Vegard's law, whereas the other alloys all showed slightly negative deviation from Vegard's law. 

\begin{figure}[!tbp]
\includegraphics[width=0.40\textwidth]{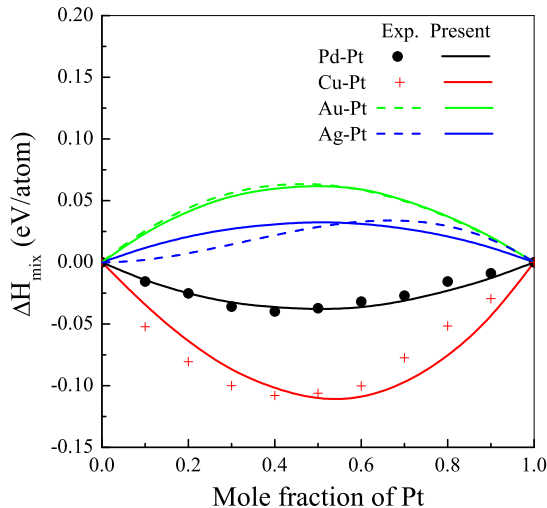}
\caption
{ \label{fig:dHmix}
Heat of formation for the Pd-Pt, Cu-Pt, Ag-Pt, and Au-Pt bulk systems as a function of Pt concentration. The solid lines are the results of the present EAM model for a disordered solid solution (random mixing). The solid symbols represent the experimental measurement of the heat of formation (Refs.~\onlinecite{Darby:ThermoPdPt} and \onlinecite{Abe:ThermoCuPt}) and the dashed lines represent the model equations that are derived from the experimental data and the phase diagrams (Refs.~\onlinecite{Karakaya:ThermoAuPt} and \onlinecite{Durussel:ThermoAgPt}).
}
\end{figure}

Figure~\ref{fig:dHmix} shows the heat of formation for Pd-Pt, Cu-Pt, Ag-Pt, and Au-Pt bulk alloys. The calculated heats of formation data for disordered solid solutions were compared with the experimental data. Our present model showed good agreement with the experimental data and the empirical model predictions. Interestingly, the Pd-Pt and Cu-Pt systems showed negative heat of formation behavior, whereas the Au-Pt and Ag-Pt system showed positive heat of solution behavior in the bulk phase.

\section{\label{sec:level_3method} Computational Method}

\subsection{\label{sec:sublevel_31MC} Initial Structures of the Nanoparticles}

\begin{figure}[!tbp]
\includegraphics[width=0.48\textwidth]{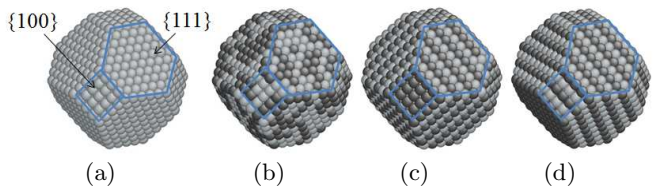}
(a) \hspace*{1.60cm} (b) \hspace*{1.30cm} (c) \hspace*{1.30cm} (d)
\caption
{ \label{fig:initialcfg}
Initial atomic configurations: (a) a monometallic nanoparticle of fcc structure and bimetallic nanoparticles of (b) random mixing, (c) $\rm L1_0$ and (d) $\rm L1_1$ ordered structure.
}
\end{figure}

It is well known that the metallic nanoparticles present competitive structural motifs, such as the icosahedron, decahedron, cuboctahedron, and truncated octahedron (TOh)~\cite{Nam:IcosahedronPRL, Nam:IcosahedronPRB}. In our atomistic simulations, a truncated octahedron (TOh) nanoparticle of 1654 atoms was used as an initial atomic configuration, which corresponds to $\sim$3 to 4 nm in size, depending on its lattice distance. The TOh shape of the nanoparticles was chosen because it is the experimentally observed shape for this size of fcc metal nanoparticle~\cite{Ferrando:NanoclusterRMP}. As shown in Fig.~\ref{fig:initialcfg}, the TOh shape is closed-shell structures having 6 $\{100\}$ facets, 8 $\{111\}$ facets, 12 $\{111\}$/$\{111\}$ edges, 24 $\{111\}$/$\{100\}$ edges, and 24 vertices and the outermost surface layer is composed of 582 atoms (35\%). Although we investigated the structures of various mole fractions, the composition of 50 at\% Pt was chosen as an example condition to simplify the discussion. With the composition fixed at 1:1 stoichiometry, it appears that any combinations of the core-shell structure are possible. A starting compositional configuration is typically generated in a random way according to the nominal composition of the bulk alloy [Fig.~\ref{fig:initialcfg} (b), although we also used alloy nanoparticles with some initially ordered configuration for comparison [Fig.~\ref{fig:initialcfg} (c) and (d)]. All of the atoms were fully relaxed with molecular statics. 

\subsection{\label{sec:sublevel_32MC} Monte Carlo Simulations of the Nanoparticles} 

As a first step to investigate the structural behavior of alloy nanoparticles, we focused on the energetics. For nanoparticles, the prediction of energetically stable structures is regarded as an optimization problem of high complexity because it is a problem of searching for the lowest energy configuration in the solution space composed of all feasible atom arrangements. Although it is not perfect for global optimization, Monte Carlo (MC) simulation is the most popular method for this purpose~\cite{Frenkel:UnderstandingMS}. In this study, we performed a series of Monte Carlo simulations based on the Metropolis algorithm, where not only the number of total atoms but also that of each element in the alloy is kept constant: The zero-temperature minimum-energy configurations of alloy nanoparticles were predicted by optimizing their total energies with respect to their compositional exchange. Because the prediction quality of a Monte Carlo simulation might depend on the initial condition, we adopted various initial configurations, including a random mixture and some ordered structures, as shown in Fig.~\ref{fig:initialcfg}. The zero-temperature Monte Carlo simulations were performed for at least $10^5$ Monte Carlo steps (MCS) for the optimizations, where one MCS corresponds to $N$ attempted exchanges and $N$ is the number of atoms in the nanoparticle. For the simulations, we used a hybrid scheme where structural relaxation using the conjugate-gradient method was implemented in the trials of compositional exchange~\cite{Frenkel:UnderstandingMS}. To avoid being trapped in a local minimum, we also performed a series of simulated annealing (SA) optimizations combined with the Monte Carlo simulations~\cite{Kirkpatrick:SimulatedAnnealing}. Here, the simulated annealing procedure was performed from T = 1000 K to 0 K at a cooling rate of of $\sim 10^{-2}$ K/MCS.

\section{\label{sec:level_4Result} Results and Discussion}

\subsection{\label{sec:sublevel_41EnergeticsTOh} Structure of TOh Alloy Nanoparticles}

\begin{figure}[!tbp]
\includegraphics[width=0.40\textwidth]{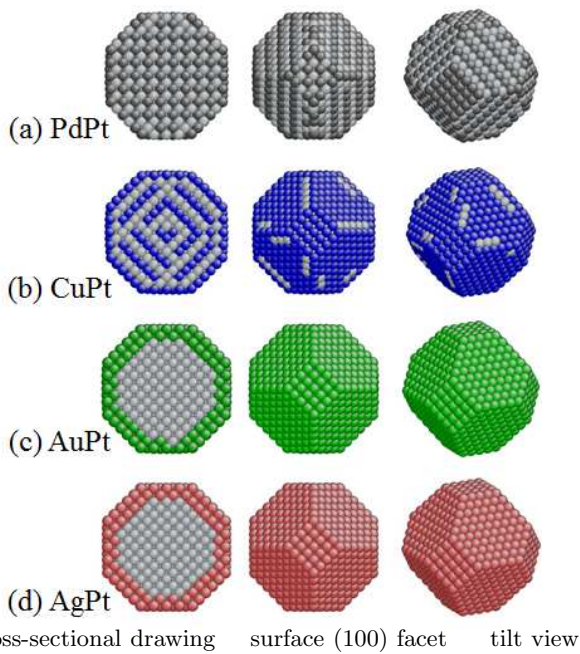}
cross-sectional drawing \hspace*{0.25cm} surface $(100)$ facet \hspace*{0.35cm} tilt view 
\caption
{ \label{fig:optimizedcfgTOh}
Predicted atomic structures of (a) Pd-Pt, (b) Cu-Pt, (c) Au-Pt, and (d) Ag-Pt nanoparticles.
The first, middle, and last columns of figures show the cross-section, surface, and tilt view of the nanoparticles, respectively.
The TOh nanoparticles are composed of 1654 atoms (M$_{827}$Pt$_{827}$, where M stands for Ag, Au, Cu, and Pd).
Pt atoms are in grey (light), Pd in black, Cu in blue, Au in green, and Ag in red (dark).  
}
\end{figure}

Although numerical variation was present in the final binding energy of the optimized structures, both the simulated annealing and the simple Monte Carlo simulations produced nearly identical views of the bimetallic nanoparticle structures. Also, the optimized structures were relatively insensitive to the initial structures, probably because of the low transition barrier for compositional configuration change from a random structure to the optimized structure. Figure~\ref{fig:optimizedcfgTOh} shows the lowest energy configurations of the 1654 atoms of TOh nanoparticles for the four alloy systems. Of course, the bimetallic nanoparticles show different compositional configurations depending on the component metals: (1) The Pd-Pt nanoparticle exhibits a typical $\rm L1_0$ intermetallic compound structure with weak surface segregation; (2) The Cu-Pt nanoparticle exhibits a onion-like multi-shell structure with Cu on the outer surfaces; (3) The Au-Pt and Ag-Pt nanoparticles exhibit Au@Pt and Ag@Pt core-shell structures as minimum energy states. 

\begin{figure}[!tbp]
\includegraphics[width=0.40\textwidth]{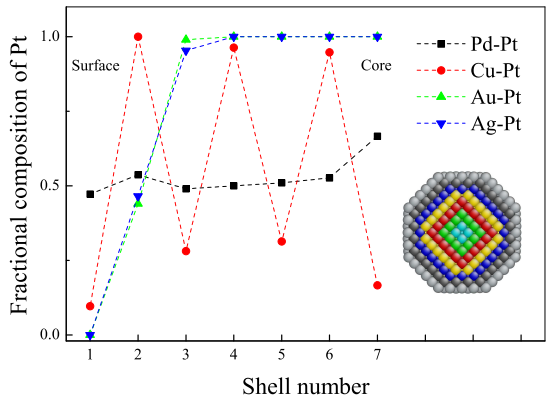}
\caption
{ \label{fig:shellprofileTOh}
Pt compositional profile in the shells of the 1654 atom TOh alloy nanoparticles. The inset shows seven shells of the nanoparticle in different colors: the first shell number denotes the surface, the second shell denotes the subsurface, and so on. 
}
\end{figure}

The inner structure can be analyzed more clearly in Fig.~\ref{fig:shellprofileTOh}, where the composition profiles of Pt in each of the shells of the nanoparticles were plotted. In the Pd-Pt nanoparticle, Pd and Pt atoms are comparably distributed over the shells, indicating ordered structures. The Cu-Pt nanoparticle exhibits an alternating distribution of Cu and Pt atoms over the shells: The 1st, 3rd, and 5th shells are Cu-rich, whereas the 2nd, 4th, and 6th shells are exclusively occupied by Pt atoms. Note that after peeling off the Cu-rich outer shell, we can obtain a perfect Pt surface. In Au-Pt and Ag-Pt nanoparticles, Pt atoms exist only on the core region and are completely surrounded by the Au or Ag shell (Au@Pt and Ag@Pt core-shell nanoparticles). 

\subsection{\label{sec:sublevel_42BulkAlloyPhase} Equilibrium Structure of Bulk Alloy Phases}

\begin{figure}[!tbp]
\includegraphics[width=0.40\textwidth]{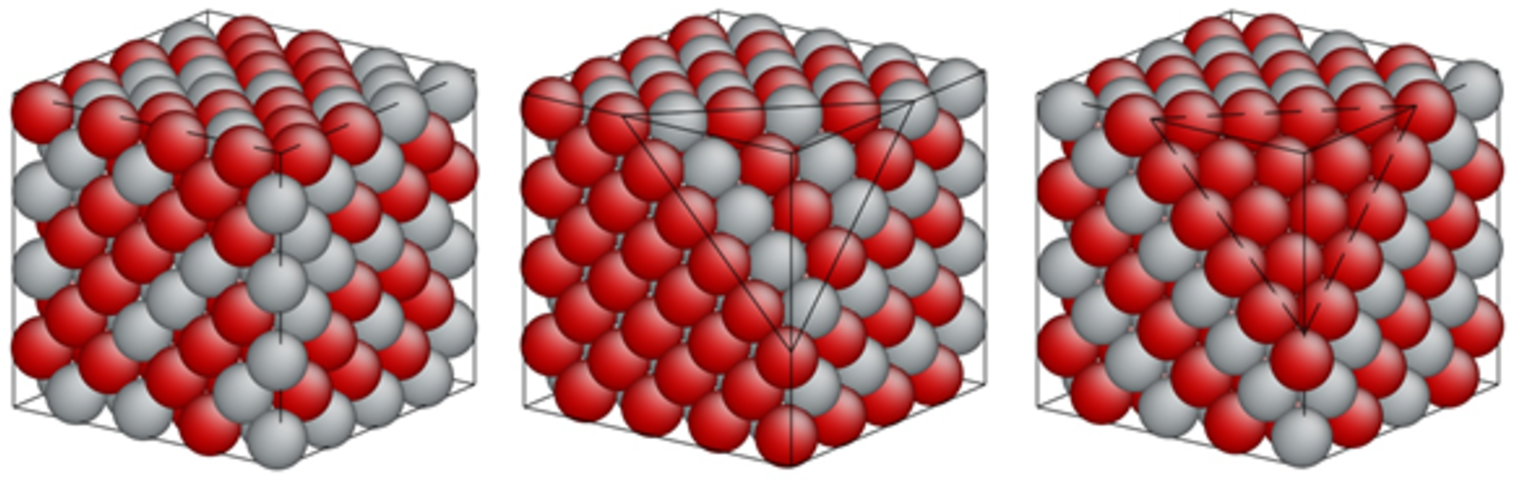}

(a) solid solution \hspace*{0.30cm} (b) $\rm L1_0$ \hspace*{1.00cm} (c) $\rm L1_1$

\includegraphics[width=0.40\textwidth]{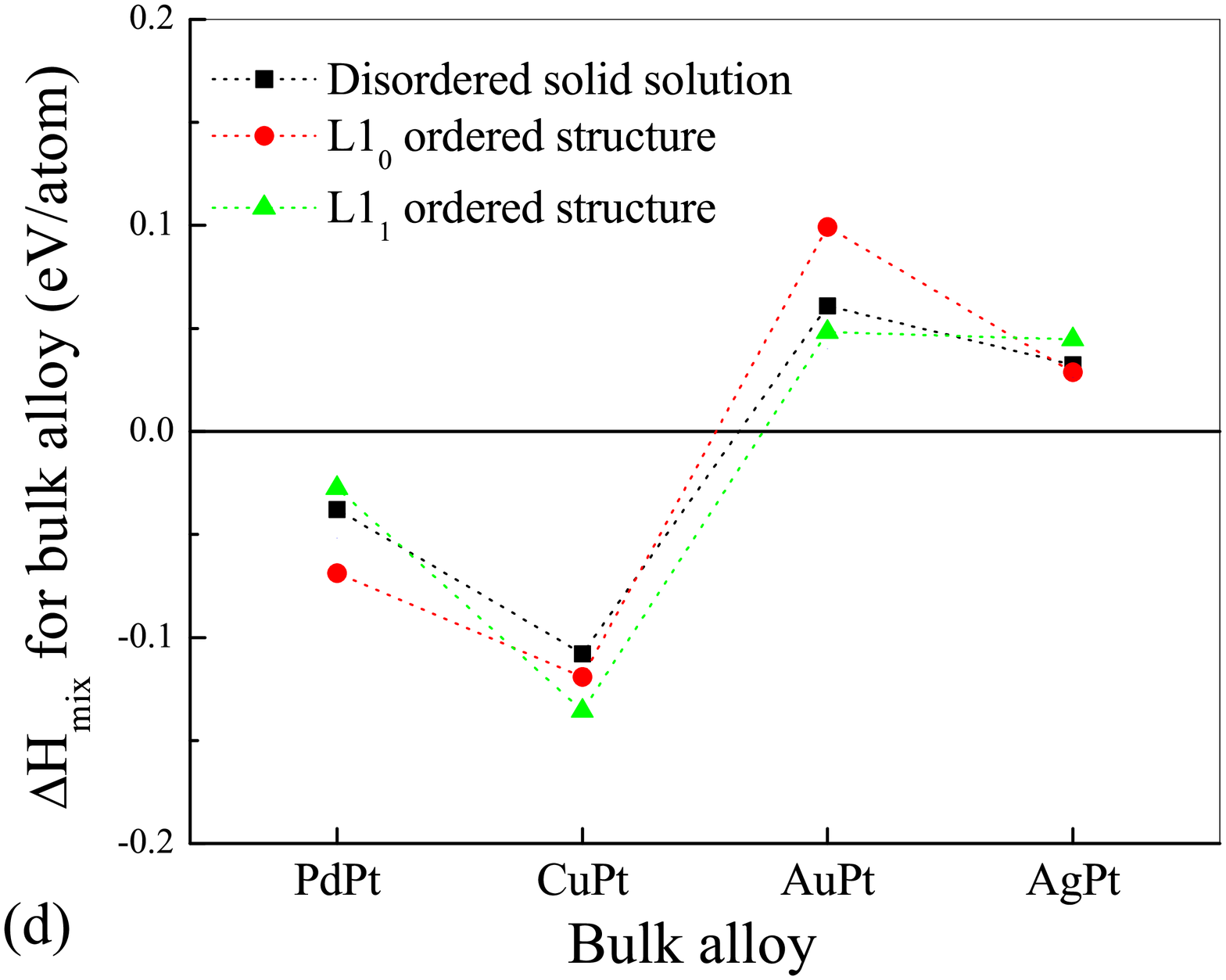}

\caption
{ \label{fig:dHformalloy}
Three types of binary alloy structures: (a) disordered solid solution, (b) $\rm L1_0$, and (c) $\rm L1_1$ crystal lattice. 
(d) Heats of formation for the three alloy crystal types in the Pd-Pt, Cu-Pt, Au-Pt, and Ag-Pt alloy systems.
}
\end{figure}

Some material properties that seem to be deeply related to alloy structures are already shown in Fig.~\ref{fig:SurfaceE} (surface energy), Fig.~\ref{fig:MeltingPt} (cohesive energy and melting point), and Fig.~\ref{fig:dHmix} (heat of random mixing) in Set.~\ref{sec:level_2potential}. In addition, the calculated heats of formation for three different structures of M$_{50}$Pt$_{50}$ (M 50\% - Pt 50\%, where M stands for Ag, Au, Cu, and Pd) bulk alloys, including one disordered and two ordered phases, are given in Fig.~\ref{fig:dHformalloy}. As previously shown in Fig.~\ref{fig:dHmix}, Pd and Cu exhibit negative heats of mixing with Pt, whereas Au and Ag exhibit positive heats of mixing behavior with Pt in the bulk phases. At the particular composition of 1:1, the tendency for ordering is strong in both Pd-Pt and Cu-Pt alloys~\cite{Johnson:OrderingPtAlloys}. For example, our potential model predicts that the Pd-Pt bulk alloy clearly favors the intermetallic compound phase of the $\rm L1_0$ ordered structure as a minimum energy state. The Cu-Pt alloy shows even larger negative heat of mixing and Cu$_{50}$Pt$_{50}$ favors the intermetallic compound phase of the $\rm L1_1$ ordered structure over random mixing or the $\rm L1_0$ structure. Interestingly, CuPt is known as the only metallic alloy system that forms in the $\rm L1_1$ structure~\cite{Massalski:BinaryAlloyPhaseDiagrams}. This structure consists of alternating fcc $(111)$ layers of Cu and Pt, in contrast with the $\rm L1_0$ structure of alternating $(001)$ planes of atoms, as shown in Figs.~\ref{fig:dHformalloy} (b) and (c). 

Our EAM alloy model predicts that the orders of formation energy for the three bulk alloy structures are  $\Delta H (\rm L1_1)$ $> \Delta H$(disordered) $> \Delta H (\rm L1_0)$ for Pd-Pt and $\Delta H$ (disordered) $> \Delta H (\rm L1_0)$ $> \Delta H (\rm L1_1)$ for Cu-Pt, respectively.  Note that the predicted lowest energy crystal structures for Pd$_{50}$Pt$_{50}$ and Cu$_{50}$Pt$_{50}$ bulk alloys are in excellent agreement with the experiments~\cite{Darby:ThermoPdPt, Abe:ThermoCuPt}. In contrast, the formation energy of mixed structures is less important for Au-Pt and Ag-Pt, because none of them is an equilibrium structure over the separated phase. Regardless, because the bulk structure is deeply related to the inner structure of the nanoparticles, accurate reproduction of the bulk alloy structures is critical for reasonable prediction of the optimized structures of the Pd-Pt and Cu-Pt nanoparticles. 

\subsection{\label{sec:sublevel_43PdPt_CuPt} Structural features of Pd-Pt and Cu-Pt nanoparticles}

As depicted in Fig.~\ref{fig:optimizedcfgTOh}, the tendency for ordering is strong in both the Pd-Pt and Cu-Pt alloy nanoparticles. The Pd-Pt nanoparticle forms a mixed alloy of ordered structure, as does the bulk PdPt alloy. However, the Cu-Pt nanoparticle forms a unique crystal structure that cannot be observed in bulk structures: it forms an onion-like structure (with surface segregation of Cu) rather than the $\rm L1_0$ or $\rm L1_1$ structures. To analyze the bond characteristics of the alloys, we analyzed the structural properties with the Warren-Cowley chemical short-range order (CSRO) parameter defined as~\cite{Cowley:CSROPRB}

\begin{equation}
\label{eq:CSRO}
{\rm CSRO} = 1- \frac {N_{AB}} {N \cdot C_B},
\end{equation}
where $N_{AB}$ is the nearest coordination number of B atoms around an A atom, N is the total coordination number in the nearest-neighbor shell, and $C_B$ is the atomic concentration of B. 

\begin{figure}[!tbp]
\includegraphics[width=0.30\textwidth]{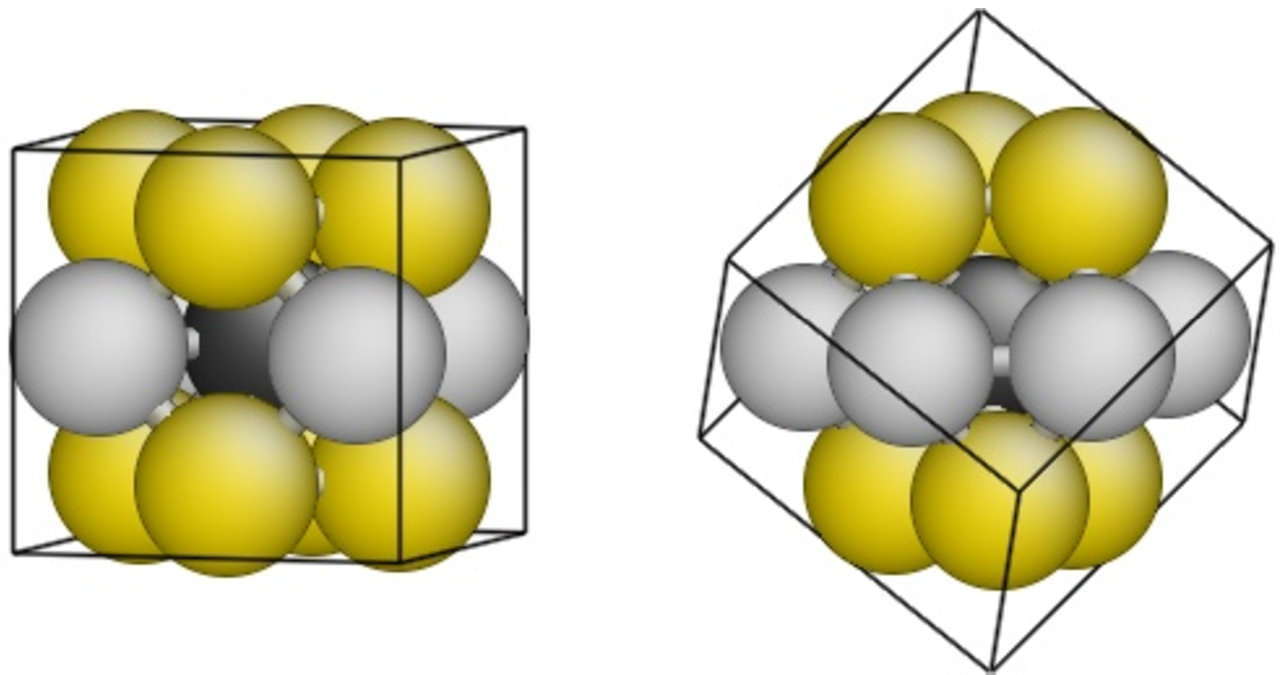}

(a) $\rm L1_0$ \hspace*{1.00cm} (b) $\rm L1_1$ 

\includegraphics[width=0.40\textwidth]{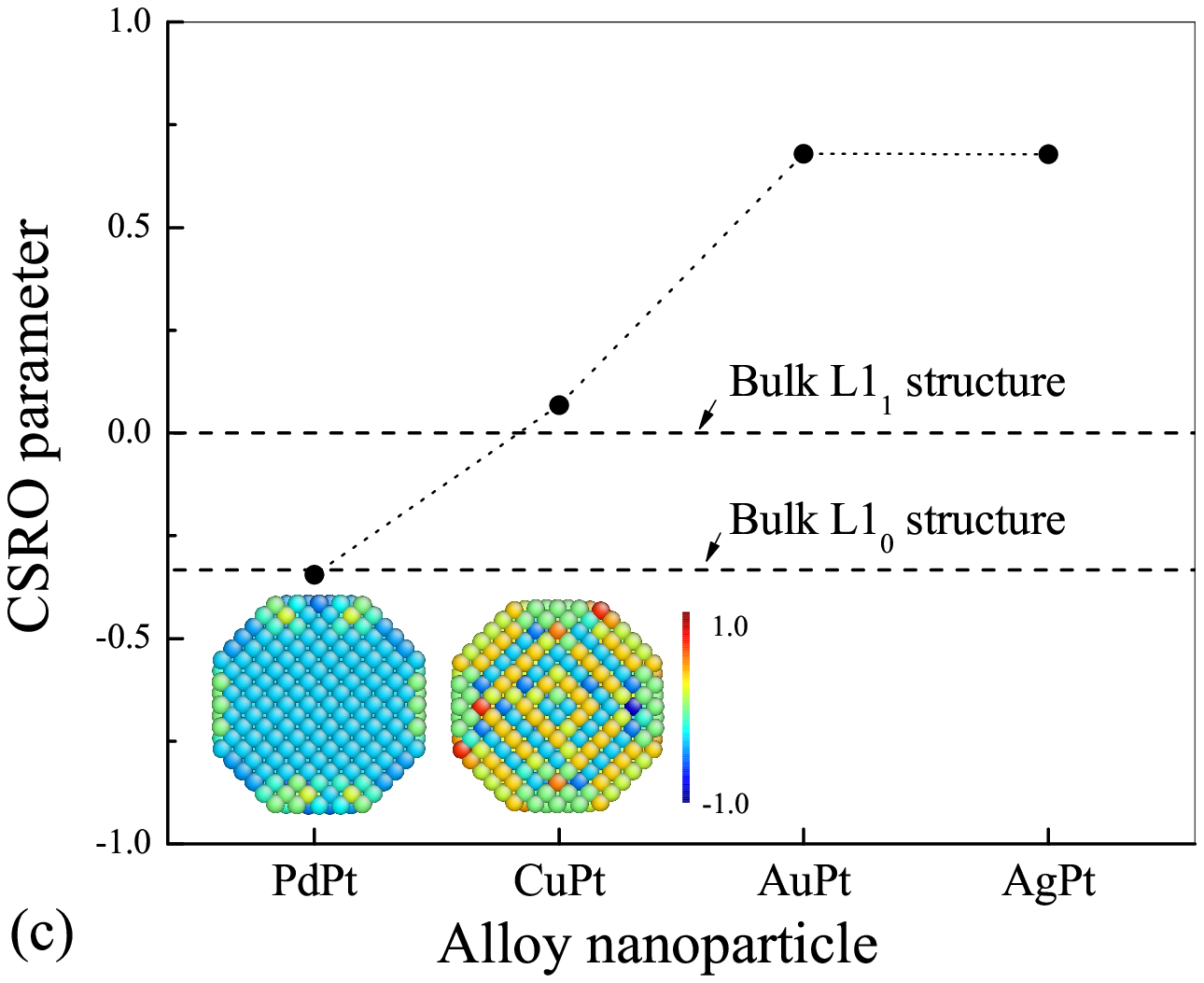}
\caption
{ \label{fig:CSROTOh}
Local atomic configurations in (a) $\rm L1_0$ and (b) $\rm L1_1$ crystal lattices; (c) Warren-Cowley chemical short-range order (CSRO) parameters for the four alloy nanoparticles.
}
\end{figure}

This parameter is a useful measure of the chemical affinity and represents the degree of tendency for ordering or clustering: CSRO values range between -1 to 1 and positive CSRO parameter values indicate clustering or phase separation and negative values indicate strong A-B bond and chemical ordering. As depicted in Figs.~\ref{fig:dHformalloy}(b) and (c), the $\rm L1_0$ structure consists of alternating fcc $(001)$ planes of Cu and Pt, whereas the $\rm L1_1$ structure consists of alternating fcc $(111)$ planes of atoms. These atomic arrangements of different alloy lattice types can also be distinguished by the CSRO values. For example, atoms in the $\rm L1_0$ crystal structure have 4 like-atom neighbors in the same $(001)$ plane and 8 (=4+4) unlike-atom neighbors in two nearby $(001)$ planes as shown in Fig.~\ref{fig:CSROTOh}(a). So the local CSRO value should be 1/3 for the $\rm L1_0$ crystal structure. Likewise, atoms in the $\rm L1_1$ crystal structure have 6 like-atom neighbors in the same $(111)$ plane and 6 (=3+3) unlike-atom neighbors in the nearby $(111)$ planes, as shown in Fig.~\ref{fig:CSROTOh}(b), and the local CSRO value goes to zero, regardless of their element type. 

Figure~\ref{fig:CSROTOh}(c) shows the CSRO parameter for the four alloy nanoparticles presented in Fig.~\ref{fig:optimizedcfgTOh}. Whereas the Au-Pt and Ag-Pt nanoparticles exhibit positive CSRO parameters with values close to 1, indicating strong separation, the CSRO of the Pd-Pt nanoparticle is approximately  $-0.33$, which is nearly the ideal value of the bulk $\rm L1_0$ structure, and the inset to the figure indicates nearly identical ordering inside the nanoparticle for the $\rm L1_0$ structure. However, the CSRO of the Cu-Pt onion-like nanoparticle is very close to zero, which is the ideal value for bulk $\rm L1_1$ structures. Although the apparent atomic configurations appear to be quite different from each other, the onion-like structure is, however, very similar to $\rm L1_1$ in terms of the chemical short-range-order (CSRO). Because they are both ordered structures of multi-layers or multi-shells, the overall chemical bond status is expected to be very similar. 

The transition of the lowest energy structure of certain alloy nanoparticles from the $\rm L1_1$ structure to an onion-like structure is very interesting, and we expect that it is probably due to surface effects. To understand the role of surface energy in nanoparticle structures, we summarized the $(001)$ and $(111)$ surface energies of various relevant Pd-Pt and Cu-Pt alloy structures that can appear in TOh nanoparticles, as shown in Table~\ref{tab:table4}. All of the surface energies were calculated based on our present EAM potentials using a slab of more than 100 atomic layers under a 2-dimensional periodic boundary condition.

\begin{table}[!tbp]
\caption
{\label{tab:table4}
Comparison of the surface energies for various surface types. The numbers in parentheses are relative values with respect to the surface energy of pure Pt. 
}
\begin{ruledtabular}
\begin{tabular}{c r@{}l r@{}l}
Bulk structure  & \multicolumn{2}{c}{Surface energy [mJ/m$^2$]} \\
                & \multicolumn{1}{c}{(001)} & \multicolumn{1}{c}{(111)} \\
\hline \\  [-1.5ex]
Pt (fcc)        & 1891 (1.00) & 1580 (1.00)       \\ [1.0ex]
Pd (fcc)        & 1815 (0.96) & 1602 (1.01) \\ [1.0ex]
Cu (fcc)        & 1435 (0.76) & 1266 (0.80) \\ [1.0ex]
PdPt (L1$_{0}$) & 1843 (0.97) & 1631 (1.03) \\ [1.0ex]
CuPt (L1$_{1}$) & 1846 (0.98) & 1967 (1.25) \\ [1.0ex]
\end{tabular}
\end{ruledtabular}
\end{table}

For Pd-Pt, both the $(100)$ and $(111)$ surface energies of Pd are comparable to those of Pt. The surface energies of the alloy $\rm L1_0$ phase, which is the equilibrium state in bulk, are also comparable to those of pure Pt or Pd (within less than a 10\% difference). The negligible dependence of surface energy on alloy composition may be attributed to the similar atomic sizes and bonding characteristics of Pd and Pt, such as cohesive energy and melting point. As shown in Fig.~\ref{fig:initialcfg}, ideal TOh nanoparticles of $\rm L1_0$ structure can have several different types of facets, including $(100)$ facets of pure Pt or Pd and $(111)$ facets of alloy composition. Because of the comparable surface energy properties between those facets, the Pd-Pt nanoparticles can maintain a similar structure to the bulk $\rm L1_0$ phase with only a small surface modification, even at this nanoscale.

For the Cu-Pt system, the $(100)$ and $(111)$ surface energies of Cu are much lower than those of Pt: the surface energy ratio of Cu with respect to that of Pt is approximately 0.76 for the $(100)$ surface and 0.80 for the $(111)$ surface. However, the surface energies of the alloy $\rm L1_1$ phase, which is the equilibrium state in bulk, are comparable or even higher than those of Pt, let alone Cu. In particular, the $(111)$ surface energy of the CuPt alloy $\rm L1_1$ phase, one of the dominant facet types in the TOh shape, is higher than that of Cu by $\sim50$\%. The remarkable increase in surface energy when forming an alloy structure may be partially attributed to the large atomic size mismatch between Pt and Cu (by approximately 10\%) that can produce surface waves. Because the surface energy tends to follow the bulk binding energy, the large negative heat of formation may also contribute to the large surface energy of the alloy composition. In any case, the surface energy difference between the two elements, which is not critical in bulk mixtures, becomes a crucial factor in nanoparticle formation: The more than 20\% difference in surface energy between Cu and Pt/Cu-Pt leads to the pure Cu $(100)$ and Cu-rich $(111)$ facets, resulting in the onion-like structures, as shown in Fig.~\ref{fig:optimizedcfgTOh}. This mechanism can operate effectively in minimizing the surface energy, especially when one of the component species can take advantage of a significantly lower surface energy. Due to the interplay of chemical ordering and the surface energy effect, we now believe that an alloy of $\rm L1_1$ ordered structure in bulk tends to favor an onion-like multi-shell structure when it forms a nanoparticle. To this end, some alloy systems have been reported to exhibit onion-like structures in previous reports of MD simulations~\cite{Baletto:Onion, Cheng:Onion}. However, we found that the structure of the alloy nanoparticle is very sensitive to the empirical potentials adopted in the simulation and care must be taken in the interpretation.

\subsection{\label{sec:sublevel_44AuPt_AgPt} Core-Shell Structures (Au@Pt and Ag@Pt) of Au-Pt and Ag-Pt nanoparticles}

Although Pt, Au, and Ag are fcc metals, and the mismatch of the lattice constants between Pt and Au or Pt and Ag is relatively small ($\sim 4$\%, see Table~\ref{tab:table2}), the Au-Pt and Ag-Pt alloy systems are thermodynamically immiscible, exhibiting large positive heat of mixing values. Referring to the phase diagrams of bulk alloy systems, we know that Ag and Pt form a peritectic system and that Ag and Pt form a miscibility gap. Also, the phase diagrams indicate that the mutual solubility of the Au-Pt and Ag-Pt system is almost zero or less than 4\% at room temperature~\cite{Massalski:BinaryAlloyPhaseDiagrams}. Therefore, a bulk solid alloy of 1:1 composition would decompose into the fcc fcc $\alpha_1$ and $\alpha_2$ phases of limited solid solubility. Regarding the formation of nanoparticles, our simulations predict that the Au-Pt and Ag-Pt alloy nanoparticles energetically favor Pt-core-Ag-shell and Pt-core-Au-shell atomic structures. In addition to the phase separation tendency due to the positive heat of mixing, the formation of the core-shell structure can be interpreted by the surface energy effect. As shown in Fig.~\ref{fig:SurfaceE}, the surface energies of both Au and Ag are only approximately 50 to $\sim 60$\% of that of Pt. Because of the differences in surface energy between Pt and Au/Ag, Pt always forms the core, whereas Au or Ag forms an outer shell to minimize the surface energy. Our prediction of core-shell structures (Au@Pt and Ag@Pt) for Au-Pt and Ag-Pt nanoparticles agrees well with the first-principle calculation~\cite{Johnston:NPsAuPtDFT} and Monte Carlo studies~\cite{Deng:NPsAuPt_SurfaceSegregation}, although the size and shape conditions are slightly different.

\section{\label{sec:level_5Conclusion} Conclusion}

For systematic design of bimetallic nanoparticle catalysts, we developed a unified EAM potential model for various Pt-based binary alloys, such as Pd-Pt, Cu-Pt, Au-Pt, and Ag-Pt. After verifying that our alloy potentials are reliable enough to accurately reproduce the surface energy and heat of formation of alloy mixtures, we performed a series of simulated annealing Monte Carlo optimizations on TOh-shaped bimetallic nanoparticles consisting of 1654 atoms. Our simulations quantify the energetically favorable atomic arrangements of the four Pt-based alloy nanoparticles: the $\rm L1_0$ intermetallic compound structure for the Pd-Pt system, an onion-like multi-shell structure for the Cu-Pt system, and core-shell structures (Au@Pt and Ag@Pt) for the Au-Pt and Ag-Pt systems. The main features of these optimized nanoparticle structures were analyzed on the basis of the relationship with their bulk alloy structures. The equilibrium structure of bimetallic nanoparticles could be interpreted by the interplay of material properties, such as the surface energy and heat of formation. We believe that understanding the energetically favorable structures can be a first step to the new design of nanoparticle structures. To this end, the relationship between the material properties and the resulting nanoparticle structures is of primary importance; other parameters, such as the size, composition, and temperature effect will be investigated in future studies.


\begin{acknowledgments}
This work was partly supported by the ERC program (R11-2005-048-00000-0), the Priority Research Centers Program (2009-0093814), and the Nano R\&D program (2011-0019162) through the National Research Foundation (NRF) funded by the Ministry of Education, Science and Technology (MEST) of Korea. S.C.L. acknowledges support from Future Research program (2E22121) funded by Korea Institute of Science and Technology. H.-S.N. acknowledges the support of the faculty research program 2010 of Kookmin University in Korea.
\end{acknowledgments}



\bibliography{PtAlloyNPs_arXiv}

\end{document}